	\documentclass[floatfix,aps,twocolumn,superscriptaddress]{revtex4-1}

\usepackage{amssymb}
\usepackage{amsmath}
\usepackage[dvips]{graphicx}
\graphicspath{{./figures/}}
\usepackage{bm}
\usepackage{float}
\usepackage{color}
\usepackage{mathtools}

\newcommand{\be}{\begin{equation}}
\newcommand{\ee}{\end{equation}}
\newcommand{\bea}{\begin{eqnarray}}
\newcommand{\eea}{\end{eqnarray}}

\begin{document}

\title{SU(4) Kondo entanglement in double quantum dot devices}

\author{Rodrigo Bonazzola}
\affiliation{Centro At{\'o}mico Bariloche and Instituto Balseiro, CNEA, 8400 Bariloche, Argentina}
\author{J. A. Andrade}
\affiliation{Centro At{\'o}mico Bariloche and Instituto Balseiro, CNEA, 8400 Bariloche, Argentina}
\affiliation{Consejo Nacional de Investigaciones Cient\'{\i}ficas y T\'ecnicas (CONICET), Argentina}
\author{Jorge I. Facio}
\affiliation{Centro At{\'o}mico Bariloche and Instituto Balseiro, CNEA, 8400 Bariloche, Argentina}
\affiliation{Consejo Nacional de Investigaciones Cient\'{\i}ficas y T\'ecnicas (CONICET), Argentina}
\author{D. J.  Garc\'ia}
\affiliation{Centro At{\'o}mico Bariloche and Instituto Balseiro, CNEA, 8400 Bariloche, Argentina}
\affiliation{Consejo Nacional de Investigaciones Cient\'{\i}ficas y T\'ecnicas (CONICET), Argentina}
\author{Pablo S. Cornaglia}
\affiliation{Centro At{\'o}mico Bariloche and Instituto Balseiro, CNEA, 8400 Bariloche, Argentina}
\affiliation{Consejo Nacional de Investigaciones Cient\'{\i}ficas y T\'ecnicas (CONICET), Argentina}

\begin{abstract}
We analyze, from a quantum information theory perspective, the possibility of realizing an SU(4) entangled Kondo regime in semiconductor double quantum dot devices. We focus our analysis on the ground state properties and consider the general experimental situation where the coupling parameters of the two quantum dots differ. We model each quantum dot with an Anderson type Hamiltonian including an interdot Coulomb repulsion and tunnel couplings for each quantum dot to independent fermionic baths.  We find that the spin and pseudospin entanglements can be made equal, and the SU(4) symmetry recovered, if the gate voltages are chosen in such a way that the average charge occupancies of the two quantum dots are equal, and the double occupancy on the double quantum dot is suppressed. 
We present density matrix renormalization group numerical results for the spin and pseudospin entanglement entropies, and analytical results for a simplified model that captures the main physics of the problem. 
\end{abstract}


\maketitle
\section{Introduction}
Quantum information theory has proved to be a powerful tool to analyze many-body problems in condensed matter physics, both providing new insights into strongly correlated states and in the development of numerical tools \cite{amico2008entanglement,verstraete2004density,verstraete2008matrix}. Entanglement measures have been used to characterize quantum phase transitions \cite{osterloh2002scaling,osborne2002entanglement,gu2004entanglement,bayat2014order} and to analyze spatial correlations in the Kondo problem \cite{sorensen2007impurity,bayat2010negativity,bayat2012entanglement,alkurtass2016entanglement,Laflorencie20161,bayat2017scaling,feiguin2017}, a paradigm example of many-body physics. 

In the Kondo problem a spin-1/2 magnetic impurity couples antiferromagnetically to a non-interacting Fermi sea. As the temperature is lowered there is a crossover, at a characteristic temperature $T_K$, from a free magnetic moment regime to a fully screened regime. The Kondo temperature $T_K$ is the only relevant scale at low energies and all physical properties are universal functions when properly scaled by $T_K$. The Kondo problem was first encountered for atomic impurities in metallic hosts in the 1930s, remained unsolved for more than 40 years and enjoyed a revival after 1998 when it was realized in quantum dot and in single molecule devices \cite{kouwenhoven2001revival}. The development of Dynamical Mean Field Theory as a tool to study the physics of strongly correlated electron systems in the lattice spurred further interest on the Kondo problem and other related multiorbital quantum impurity problems \cite{georges2015beauty}.

These and other related models present a rich variety of phenomena as singular and non-Fermi liquid behavior and a high sensitivity to external fields and may provide a road for the understanding of strongly correlated materials. 
In particular, the SU(4) Kondo model, which has been realized in carbon nanotube devices \cite{makarovski20072,choi20054,tettamanzi2012magnetic,busser2007numerical,cleuziou2013interplay}, has an enhanced Kondo temperature and peculiar low energy spectral properties that lead to a large thermoelectric power at low temperatures \cite{roura2012thermopower,lim2014orbital}. It has also been proposed as a means to build high transmittance spin filtering devices \cite{borda2003su4,feinberg2004splitting,busser2012Electrostatic,vernek2014spin}.

The double quantum dot (DQD) device proposed in Ref. \cite{borda2003su4} to observe the SU(4) Kondo effect was realized in Ref. \cite{keller2014emergent} and an excellent agreement in the transport properties with an Anderson-like Hamiltonian was found. A scaling of the conductance consistent with an SU(4) behavior was also reported. 
Nishikawa {et al.} in Ref. \cite{nishikawa2013emergent} showed however, using a renormalized perturbation theory treatment of the Anderson model, that the low energy effective Hamiltonian is not SU(4)-symmetric unless the original model is SU(4)-symmetric or the local interactions on the DQD are very large \cite{nishikawa2013emergent,tosi2015restoring}. These conditions are not met in the experiments of Ref. \cite{keller2014emergent}.  

In this article we address this apparent contradiction using quantum information theory tools.
By analysing the symmetry of the entanglement of the ground state wave function we show that charge fluctuations into doubly occupied states must be vanishingly small or all doubly occupied states equally probable in order not to break the SU(4) symmetry. Even for SU(4)-symmetric interactions and when the charge symmetry between the quantum dots (QDs) is restored, the fluctuations to doubly occupied states break the SU(4)-symmetry of the ground state wave function when the electrode-QD hybridizations of the two QDs differ. We show that it is possible to reduce the size of SU(4) symmetry breaking in the entanglement by reducing the DQD average occupancy to less that one electron, therefore suppressing charge fluctuations to the doubly occupied states. Interestingly, the results for the SU(4) scaling presented by Keller {\it et al.} (Ref \cite{keller2014emergent}) are precisely in a regime with less than a single electron on average on the DQD.

The rest of the paper is organized as follows. The model Hamiltonian and the methods are presented in Sec. \ref{sec:model}. Numerical results for the entanglement entropy together with analytical results for a toy model are presented in Sec. \ref{sec:results}. Finally, a summary and the conclusions are presented in Sec. \ref{sec:concl}.

\section{Model and Methods}\label{sec:model}
\begin{figure}[t]
\includegraphics[width=0.8\columnwidth]{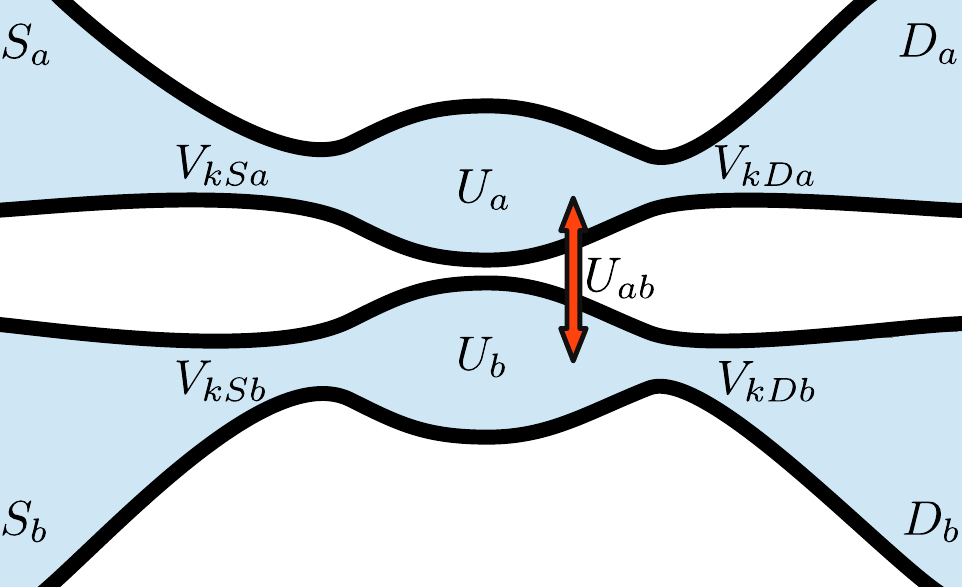}
\caption{(Color online) Schematic representation of the double quantum dot device. Each quantum dot ($a$ and $b$) is tunnel coupled to a pair of source ($S_a$ and $S_b$) and drain ($D_a$ and $D_b$) electrodes. The local ($U_a$ and $U_b$) and the interdot ($U_{ab}$) electron-electron repulsion, and the tunnel coupling of the quantum dots to the electrodes ($V_{k\nu\ell}$ with $\nu=S,D$, and $\ell=a,b$) are indicated in the figure.}
\label{fig:device}
\end{figure}
We consider a double quantum dot device DQD (see Fig. \ref{fig:device}), with a single relevant electronic level on each QD. The device is described by the following Hamiltonian
\begin{equation} \label{eq:hamilt}
  H=H_{\text{C}}+ H_{\text{V}}+H_{\text {el}}\;.
\end{equation}
Here $H_C$ describes the electrostatic interaction on the QDs 
\begin{equation} \label{eq:coul}
	H_C = \sum_{\ell=a,b}\left(\sum_{\sigma=\uparrow,\downarrow}\varepsilon_{\ell\sigma}\hat{n}_{\ell\sigma}+ U_\ell \hat{n}_{\ell\uparrow}\hat{n}_{\ell\downarrow}\right) + U_{ab}\hat{n}_{a}\hat{n}_{b},
\end{equation}
where $\sigma=\pm 1/2$ indicates the electron spin projection along the $\hat{z}$ axis,  $\hat{n}_{\ell\sigma}= d_{\ell\sigma}^\dagger d_{\ell\sigma}$ is the electron number operator of the $\ell$-th QD, $\hat{n}_{\ell}=\hat{n}_{\ell\uparrow}+\hat{n}_{\ell\downarrow}$,  $U_\ell$ is its charging energy,  $\varepsilon_{\ell\sigma}=\varepsilon_{\ell} + \sigma g\mu_B B$, where $B$ is magnetic field Zeeman coupled to the QDs' spins, and $\varepsilon_{\ell}=U_{\ell}-C_{g\ell} V_{g\ell}$ is controlled by a gate voltage $V_{g\ell}$, 
where $C_{g\ell}$ is the capacitance of QD $\ell$ with its corresponding gate electrode. 

\begin{equation} \label{eq:hybr}
	H_{\text{V}} = \sum_{\nu=S,D}\sum_{\ell=a,b}\sum_{k,\sigma}\left(V_{k\nu\ell}d_{\ell\sigma}^{\dagger}c_{\nu\ell k\sigma}^{}+ \text{H.c.} \right),
\end{equation}
describes the coupling between the QDs and their respective source ($S$) and drain ($D$) electrodes, 
which are modeled by non-interacting Fermi gases:

\begin{equation}
H_{\text el} = \sum_{\nu,\ell, k,\sigma} \epsilon_{\nu\ell k} c_{\nu\ell k\sigma}^\dagger c_{\nu\ell k\sigma}^{}.
\end{equation}
This model was used in Ref. \cite{keller2014emergent} and an excellent agreement with the experimental results was obtained for the conductance and low energy spectral properties.

Each QD couples to a specific combination of states from the source and drain electrodes $c_{0,\ell\sigma}=\frac{1}{V_\ell}\sum_k \sum_\nu V_{k\nu\ell} c_{\nu\ell k\sigma}$, where $V_\ell=\sqrt{\sum_k \sum_\nu |V_{k\nu\ell}|^2}$. For the study of equilibrium properties, the coupling of the QDs to the electrodes can be described using the hybridization functions $\Gamma_\ell(\omega)=\pi V_\ell^2 \rho_\ell(\omega)$, where $\rho_\ell(\omega)$ is the local density of states of the electrodes associated with the operator $c_{0,\ell\sigma}^{}$. 
For simplicity we will consider all asymmetries in the coupling to the electrodes to be described by the parameters $V_\ell$, and take $\rho_\ell(\omega) = \rho(\omega)=\frac{2}{\pi D^2}\sqrt{D^2-\omega^2}$ with support in the range $[-D,D]$.

\begin{figure}[t]
\includegraphics[width=\columnwidth]{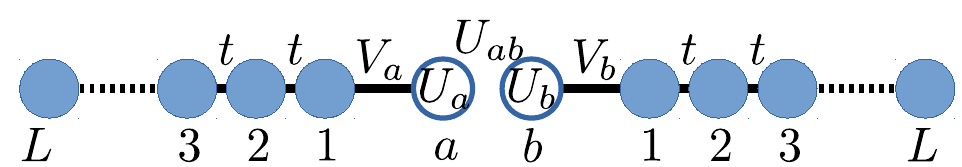}
\caption{(Color online) Schematic representation of the DQD model solved numerically using DMRG for $L$ values of up to a few hundred. The $L=1$ case can be solved analytically in the large interaction limit ($U_a, U_b,U_{ab}\to \infty$). }
\label{fig:chain}
\end{figure}

For the numerical calculations the electrodes are described using two one-dimensional tight-binding chains (one for each QD)
\begin{equation}
	H_{\text{el}} = - t \sum_{i=1}^L\sum_{\ell}\sum_\sigma \left(c_{i,\ell\sigma}^\dagger c_{i+1,\ell\sigma}^{} + \text{H.c.}\right),
	\label{eq:tbchain}
\end{equation}	
which leads to the required local density of states at site $1$ for $L\to \infty$ and $t=D/2$. The tunnel coupling is given by
\begin{equation}
	H_{\text{V}} = \sum_{\ell}\sum_{\sigma} \left(V_{\ell} d_{\ell\sigma}^{\dagger} c_{0,\ell\sigma}^{} + \text{H.c.}\right). 
	\label{eq:hybchain}
\end{equation}

This model (see Fig. \ref{fig:chain} for an schematic description) can be solved for finite $L$ using the density matrix renormalization group (DMRG) \cite{peschel1999density,schollwock2005density}. To avoid finite size effects, we performed a finite size analysis for total systems sizes $N=2L+2$ ranging from $4$ to $384$.

The case $L=1$ including a single site on each bath can be solved exactly in specific cases and will guide us in the interpretation of the numerical results.

\subsection{Low energy Hamiltonian and SU(4)-symmetry}
For symmetric parameters: $\varepsilon_a=\varepsilon_b=\varepsilon$, $V_a=V_b=V$, and $U_a=U_b=U_{ab}=U$, the Hamiltonian of the system can be written as
\begin{eqnarray}
  H_{\text{sym}} &=& \varepsilon \hat{Q} + \frac{U}{2}\hat{Q}(\hat{Q}-1) + V \sum_{\alpha}^{} \left( d_\alpha^\dagger c_{0,\alpha}^{}+ \text{H.c.}\right)\nonumber\\
	&&-t \sum_{\alpha }\sum_i \left(c_{i,\alpha}^\dagger c_{i+1,\alpha}^{}+ \text{H.c.}\right), 
	\label{eq:symham}
\end{eqnarray}
where $\alpha$ runs over the four combinations $\{a\uparrow, a\downarrow,b\uparrow, b\downarrow\}$, and $\hat{Q}=\sum_{\alpha}^{}\hat{n}_\alpha$ is the total charge operator of the DQD.

In the parameter regime where $-U<\varepsilon < 0$, and $|\varepsilon|,\,\varepsilon+U \gg \pi V^2 \rho(0)$ the average charge of the DQD in the ground state is $n=\langle Q\rangle_{\text{GS}}\sim 1$. In this case it is instructive to associate a pseudospin degree of freedom to the QD and to the non-interacting chain indices $a\to\Uparrow$, $b\to\Downarrow$. 
The fully symmetric model of Eq. (\ref{eq:symham}) is invariant under unitary transformations that mix spin and pseudospin degrees of freedom on the DQD and on each site of the non-interacting chain (the last term on Eq. (\ref{eq:symham})). A Schrieffer-Wolff transformation that decouples the empty and multiple occupied states on the DQD leads to an SU(4)-symmetric Kondo Hamiltonian with a single exchange coupling for the four combinations of spin and pseudospin degrees of freedom.
The SU(4)-symmetric Kondo Hamiltonian leads to an SU(4) Kondo effect leading to a symmetric entanglement of spin and pseudospin degrees of freedom.

In the usual experimental situation, however, the parameters are not symmetric. In the experiments of Ref. \cite{keller2014emergent} the interdot coupling $U_{ab}$ is an order of magnitude smaller than the intradot couplings $U_a$ and $U_b$ and the hybridizations are not symmetric. Performing a Schrieffer-Wolff transformation in the general case does not lead to an SU(4)-symmetric Kondo Hamiltonian. However, it was shown in Ref. \cite{borda2003su4}, using renormalization group arguments and numerical results, that an SU(4)-symmetric effective Kondo Hamiltonian is obtained at low energies.
The charge fluctuations that are eliminated from the Hamiltonian after a Schrieffer-Wolff transformation, do however change the low energy physics and need to be included in the analysis \cite{nishikawa2013emergent,tosi2015restoring}.

Using renormalized perturbation theory for identical QDs ($U_a=U_b$, $V_a=V_b$) Nishikawa {\it et al.} found that if the interdot Coulomb repulsion $U_{ab}$ is lower than the intradot repulsions $U_a,U_b$, the SU(4) symmetry is not obtained for the low energy quasiparticle Hamiltonian, unless these interaction parameters are larger than the bandwidth $D$ of the conduction band of the electrodes \cite{nishikawa2013emergent}. The spectral properties of a DQD device model with asymmetric QDs ($V_a\neq V_b$) were studied in Ref. \cite{tosi2015restoring} in the limit of large Coulomb repulsion $U_a,U_b, U_{ab}\to \infty$, where it was shown that the spectral densities of the two QDs could be made very similar by tunning the average charge on the QDs to be equal. These results were confirmed in Ref. \cite{nishikawa2016su4} through numerical renormalization group calculations which also indicate that $U_{ab}$ must be larger or of the order of $D$ for the symmetry to be restored.
In the experiments presented in Ref. \cite{keller2014emergent} the hybridizations of the two QD with the electrodes are different and the interdot interaction is ten times smaller than the intradot interaction which in turn is not larger than the electrode's bandwidth. 

In what follows we revisit this problem and study the possibility of observing an SU(4)-symmetric Kondo effect in a DQD device by analyzing the entanglement entropy of the ground state wave function.  

\subsection{Von Neumann entropy}
For a system described by the density matrix $\rho$ the Von Neumann entropy reads:
\begin{equation}
    S(\rho)=-\text{Tr}\left\{ \rho\log_2\rho \right\}
    \label{eq:vn}
\end{equation}
If the Hilbert space can be written as product of two subspaces $A$ and $B$: $\mathcal{H}=\mathcal{H}_A\otimes \mathcal{H}_B$, the density matrix for subsystem $A$ is given by 
\begin{equation}
    \rho_A=\text{Tr}_B\rho.
    \label{eq:partialtr}
\end{equation}
For a system in a pure state $\rho=|\Psi\rangle\langle\Psi|$ we can measure the entanglement between subsystems $A$ and $B$ using
\begin{equation}
    S_A \coloneqq S(\rho_A)\equiv S(\rho_B)
    \label{eq:EOF}
\end{equation}
In this work we analyze the ground state properties ($\rho=|\Psi_{\text{GS}}\rangle\langle\Psi_{\text{GS}}|$) and we will be mainly interested in the partitions of the total system $\mathcal{H}_\uparrow\otimes \mathcal{H}_\downarrow$ and $\mathcal{H}_\Uparrow\otimes \mathcal{H}_\Downarrow$ which define the entropies $S_\uparrow$ and $S_\Uparrow$, respectively.

A necessary condition for the ground state to be SU(4)-symmetric is that $S_\uparrow=S_\Uparrow$. In the next section we use this condition to explore where in the parameter space an SU(4)-symmetric ground state could be obtained.  

As we will see below, many important features of the entanglement entropy can be understood by analyzing a toy model having a single site on each fermionic bath, i.e. $L=1$ in Eq. (\ref{eq:tbchain}).

\subsection{Spin and pseudospin susceptibilities} \label{sec:sus}
Following Ref. \cite{nishikawa2013emergent} we define the zero-temperature spin susceptibility
\begin{equation}
\chi_s = \frac{d m_s}{d h_{s}},
\end{equation}
which measures the change in the spin polarization of the QDs in the ground state $m_s=\left\langle\sum_\ell(\hat{n}_{\ell\uparrow}-\hat{n}_{\ell\downarrow})\right\rangle/2$ when a Zeeman energy splitting $2 h_s= g\mu_B B$ is applied.
In analogy with the spin susceptibility, 
\begin{equation}
\chi_{ps} = \frac{d m_{ps}}{d h_{ps}},
\end{equation}
defines the pseudospin susceptibility, where 
$m_{ps}=\left\langle\sum_\sigma(\hat{n}_{a\sigma}-\hat{n}_{b\sigma})\right\rangle/2$, and $h_{ps}=(\varepsilon_b- \varepsilon_a)/2$.

In the numerical calculations presented below, the parameters are fixed in order to obtain an unpolarized ground state ($m_s=m_{ps}=0$) and a small enough energy splitting $\delta h_{{s}/{ps}}$ is applied, such that the response is linear~\footnote{An energy shift $\delta h_{{s}/{ps}}=0.0001D$ proved to be appropriate in the whole parameter regime studied}.

In the SU(4) Kondo regime these two susceptibilities are expected to be equal $\chi_s=\chi_{ps}=\chi$, and the low energy properties of the system to be universal functions when properly scaled by the Kondo energy $k_B T_K\propto 1/\chi$ \cite{wilson1975kondo}. When the SU(4) symmetry is broken, we may define a spin (pseudospin) Kondo energy $k_B T_K^{s(ps)}\propto 1/\chi_{s(ps)}$ and quantify the degree of symmetry breaking using the ratio $\delta_K = |T_K^{s}-T_K^{ps}|/ \text{min}(T_K^{s},T_K^{ps})\equiv|\chi_{s}-\chi_{ps}|/ \text{max}(\chi_{s},\chi_{ps})$. For $\delta_K \gtrsim 1$ we do not expect SU(4)-Kondo-like physics to be observed. For $\delta_K \ll 1$, however, there is an energy regime [e.g. the range of temperatures between $|T_K^{s}-T_K^{ps}|$  and $\text{min}(T_K^{s},T_K^{ps})$] where a single parameter SU(4)-Kondo scaling could be obtained. 

\section{Results}\label{sec:results}

We first analyze the entanglement entropies in the case where the Hamiltonian is given by Eq. (\ref{eq:symham}) and has SU(4) symmetry. The ground state wave function preserves the SU(4) symmetry and the associated entanglement entropies $S_\uparrow$ and $S_\Uparrow$ are identical.
The toy model (i.e. $L=1$ in Eq. (\ref{eq:tbchain})) can be solved analytically in the $U\to \infty$ limit. The ground state is in a charge sector with a total of $4$ electrons in the system of which $2$ electrons of opposite spin projection are on each QD-bath site combination. For $n\leq 1$ the wave function has the form 
\begin{equation}
  |\Psi_{\text{GS}}\rangle = \alpha |0\rangle_a \otimes |0\rangle_b + \frac{\beta}{\sqrt{2}}\left(|S\rangle_a \otimes |0\rangle_b +|0\rangle_a \otimes |S\rangle_b\right),
\label{eq:GSUinf}
\end{equation}
where $|0\rangle_i$ corresponds to QD $i$ devoid of electrons and its associated fermionic bath site doubly occupied (see table \ref{tab:states}), and $|S\rangle_i$ represents a singlet state between QD $i$ and the reservoir site coupled to it. 
\begin{table}[t]
  \centering
    \begin{tabular}{ | l | c|}
    \hline
    \textbf{State} & \multicolumn{1}{|c|}{\textbf{Configuration}} \\ \hline
    $|0\rangle_a$ & $|\underbrace{\uparrow\downarrow}_\text{1a},\underbrace{ - }_\text{a}\rangle$ \\ \hline
    $|S\rangle_a$ & $\frac{1}{\sqrt{2}}\left(|\uparrow,\downarrow\rangle - |\downarrow,\uparrow\rangle\right)$ \\ \hline
    $|2\rangle_a$ & $|-,\uparrow\downarrow\rangle$ \\ \hline
    \end{tabular}
    \caption{Notation for the relevant toy model states of the subchain including QD $a$ and its associated fermionic bath site (see Fig. \ref{fig:chain}). The notation is analogous for the subchain containing QD $b$.  }
    \label{tab:states}
\end{table} 
The entanglement entropy of the toy model can be written as a function of the DQD occupancy $n=\beta^2$:
\begin{equation}
	S_{\uparrow}^{\text{tm}}(n)=S_{\Uparrow}^{\text{tm}}(n)=h\left(\frac{1}{2}+\frac{1}{2}\sqrt{1-n^2}\right),
\label{:eq:toyUisinf}
\end{equation}
where $h(x)=-x\log _2x-(1-x)\log _2(1-x)$ is Shannon binary entropy.
For $n \to 0$, the DQD is empty ($|\Psi_{\text{GS}}\rangle\to|0\rangle_a\otimes |0\rangle_b$, which is separable) and the interaction is not effective creating correlations. As $n$ increases, $S$ increases monotonically to reach its maximum value $S=1$ that corresponds to a maximally entangled state ($\alpha=0$ and $\beta=1$) which is a superposition of spin singlets on the left and right electrodes. This state can also be rewritten as a superposition of two pseudospin singlets making explicit the symmetry between spin and pseudospin degrees of freedom. 

For $1\leq n\leq 2$, and large $U$, the wave function takes the form:
\begin{align}
  |\Psi_{\text{GS}}\rangle&=\beta\left(|S\rangle_a \otimes |0\rangle_b +|0\rangle_a \otimes |S\rangle_b\right) \\&+\frac{\gamma}{\sqrt{6}} \left(2|S\rangle_a \otimes |S\rangle_b + |2\rangle_a \otimes |0\rangle_b + |0\rangle_a \otimes |2\rangle_b\right) \nonumber
\label{eq:psi_gs}
\end{align}
where the second term is a symmetric combination of $6$ Fock states with double occupancy on the DQD.
It is also possible to obtain an analytic expression for the entanglement entropy in this case, but the expression is more complicated than in the $n\leq 1$ case. In the $n\sim 1$ regime we find
\begin{equation}
	S^{\text{tm}}(n\to 1) = \left\{\begin{matrix}1-\frac{1-n}{\log(2)}&\text{if } n<1\\1+\frac{2(1-n)}{3\log(2)}&\text{if } n>1\end{matrix}\right.
	\label{eq:entrn1}
\end{equation}
The asymmetry between $n<1$ and $n>1$ [$S^{\text{tm}}(1-\delta)>S^{\text{tm}}(1+\delta)$ for $\delta>0$] is due to the fact that while the state $|0\rangle_a \otimes |0\rangle_b$ is clearly separable both in the spin and pseudospin sectors, the combination of states with a double occupancy on the QD [second term in Eq. (\ref{eq:GSUinf})] is not. 

For finite values of $U$ the toy model can be solved numerically and the entropy as a function of the DQD occupancy has a behavior as a function of $n$ similar to the $U\to\infty$ case.  While the entanglement entropy for a given $n$ decreases with decreasing $U$, it continues to be asymmetric and presents a maximum near $n=1$.
\begin{figure}[t]
\includegraphics[width=\columnwidth]{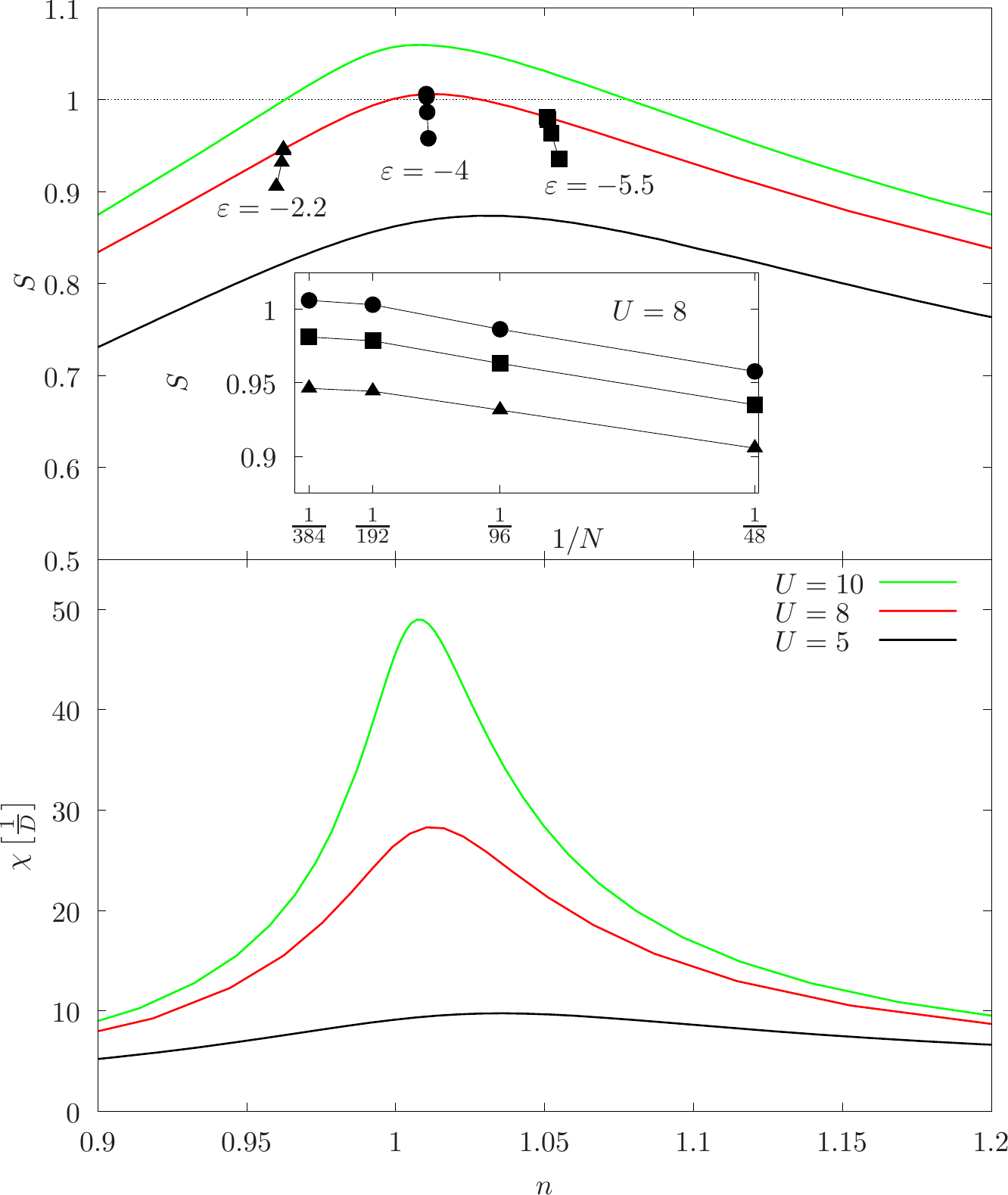}
\caption{(Color online) The upper panel presents the entanglement entropy ($S$) for systems with different values of $U=5,\,8,\, 10$, and $V_a=V_b=0.45$.  The entanglement in the spin ($S_{\uparrow}$) and pseudospin $S_{\Uparrow}$ sectors are identical for a given value of $U$. The finite size behavior of $S(n)$ for $U=8$ and a total number of DMRG calculation sites $N=2L+2$ ranging from $48$ to $384$ is presented for $\varepsilon_a=\varepsilon_b=\varepsilon=-2.2$ (solid triangles), $\varepsilon=-4$ (solid disks), and $\varepsilon=-5.5$ (solid squares).  The inset presents the behavior of $S$ as a function of $1/N$ for the values of $\varepsilon$ indicated above. The lower panel presents the spin susceptibility (which is identical to the pseudospin susceptibility) for the same set of parameters. }
\label{fig:Sym}
\end{figure}

The numerical results obtained 
using DMRG~\footnote{The DMRG results were obtained for chains of up to 384 sites, with open boundary conditions and half-filled conduction electron bands which allowed us to work in a subspace with zero total spin projection along the $z$ axis ($S^z_{tot}=0$).
The largest truncation error in the calculations, keeping up to 256 states per block and performing 6 sweeps in the finite-size algorithm, was $\rho \lesssim 10^{-5}$.
Such an error translates to an entropy defined at least with 2 decimals (relative error lower than $1\%$).} are presented in Fig. \ref{fig:Sym} as a function of the DQD occupancy $n$ for different values of the interaction $U$. The behavior of $S_\uparrow(n)$ follows qualitatively the toy model results. Namely, $S_\uparrow(n)$ has an asymmetric shape with a maximum at $n\sim 1$ and for a given value of $n$ it increases monotonically with increasing $U$. 
The most significant difference is that $S_\uparrow(n)$ becomes larger than $1$ for large enough values of $U$ and $n\sim 1$. Interestingly, these values of $S$ exceeding $1$ are due to correlations that cannot be understood using a variational wave function approach with a Varma-Yafet ansatz \cite{varma1975}. 
The spin susceptibility (see lower panel of Fig. \ref{fig:Sym}), which is identical to the pseudospin susceptibility in this fully symmetric case, presents a maximum for $n\sim 1$. As expected, the susceptibility increases for $n\sim 1$ much faster with increasing $U$ than predicted by the toy model which does not capture the Kondo renormalization of the parameters \cite{nishikawa2013emergent}.

\subsection{Broken interaction symmetry ($U_a=U_b>U_{ab}$)}
We now consider a more realistic parameter regime where the symmetry of the interaction is broken $U_{ab}<U_a,U_b$. In this case, the different states with double occupancy in the DQD are expected to have a differing participation on the GS wavefunction. In particular, the state $|S_{ab}\rangle=|S\rangle_a \otimes |S\rangle_b$, which has associated an interaction energy $U_{ab}$, is expected to have a larger amplitude in the GS wave function than the states having two electrons on a single QD (which have a larger interaction energy $U_\ell$ associated). 
This breaking of the symmetry for the amplitude in the GS wavefunction of states having a doubly occupancy in the DQD leads to a breaking of the spin-pseudospin symmetry and therefore of the SU(4) symmetry.  This effect can be understood by analyzing the structure of the ground state wave function, and the entanglement entropy of the toy model. To simplify the analysis we consider the limit $U_a,U_b\to\infty$ that captures the main features of the entropy behavior. 
The GS wave-function has the form
\begin{align}
    |\Psi_{\text{GS}}\rangle &= \alpha |0\rangle_a \otimes |0\rangle_b + \frac{\beta}{\sqrt{2}}\left(|S\rangle_a \otimes |0\rangle_b +|0\rangle_a \otimes |S\rangle_b\right)\\&+ \gamma|S\rangle_a \otimes |S\rangle_b,\nonumber
\label{eq:psi_gsU12}
\end{align}
where the double occupancy on each QD is completely suppressed and $\alpha^2+\beta^2+\gamma^2=1$.
The pseudospin entanglement entropy has the form
\begin{equation}
S_{\Uparrow}=h\bigg(\frac{1}{2}+\frac{1}{2}\sqrt{1-C^2}\bigg),
\end{equation}
where $C=|\beta^2-2\alpha\gamma|$ is Wootters' concurrence \cite{wootters1998concurrence} in the pseudospin sector. 
For $\gamma\ll 1$ (which is the case for $n<1$ and $V/U\ll 1$) we find $S_\Uparrow(n,\gamma)\leq S_\Uparrow(n,0)=S_\uparrow(n,0)\leq S_\uparrow(n,\gamma)$.

For large $U_{ab}\gg t$, $\alpha$ is strongly suppressed for $n>1$. In this regime we obtain $C=n$ for $n<1$ and $C=2-n$ for $n>1$ which leads to a mirror symmetric $S_{\Uparrow}(n)$ with respect to $n=1$. The origin of this symmetry is that the states $|0\rangle$ and $|S_{ab}\rangle$ are both separable in pseudospin. The spin entanglement entropy, however, increases monotonically with $n$ for $n>1$ and can be roughly approximated by $S_\uparrow(n>1)\sim n$.

\begin{figure}[t]
\includegraphics[width=\columnwidth]{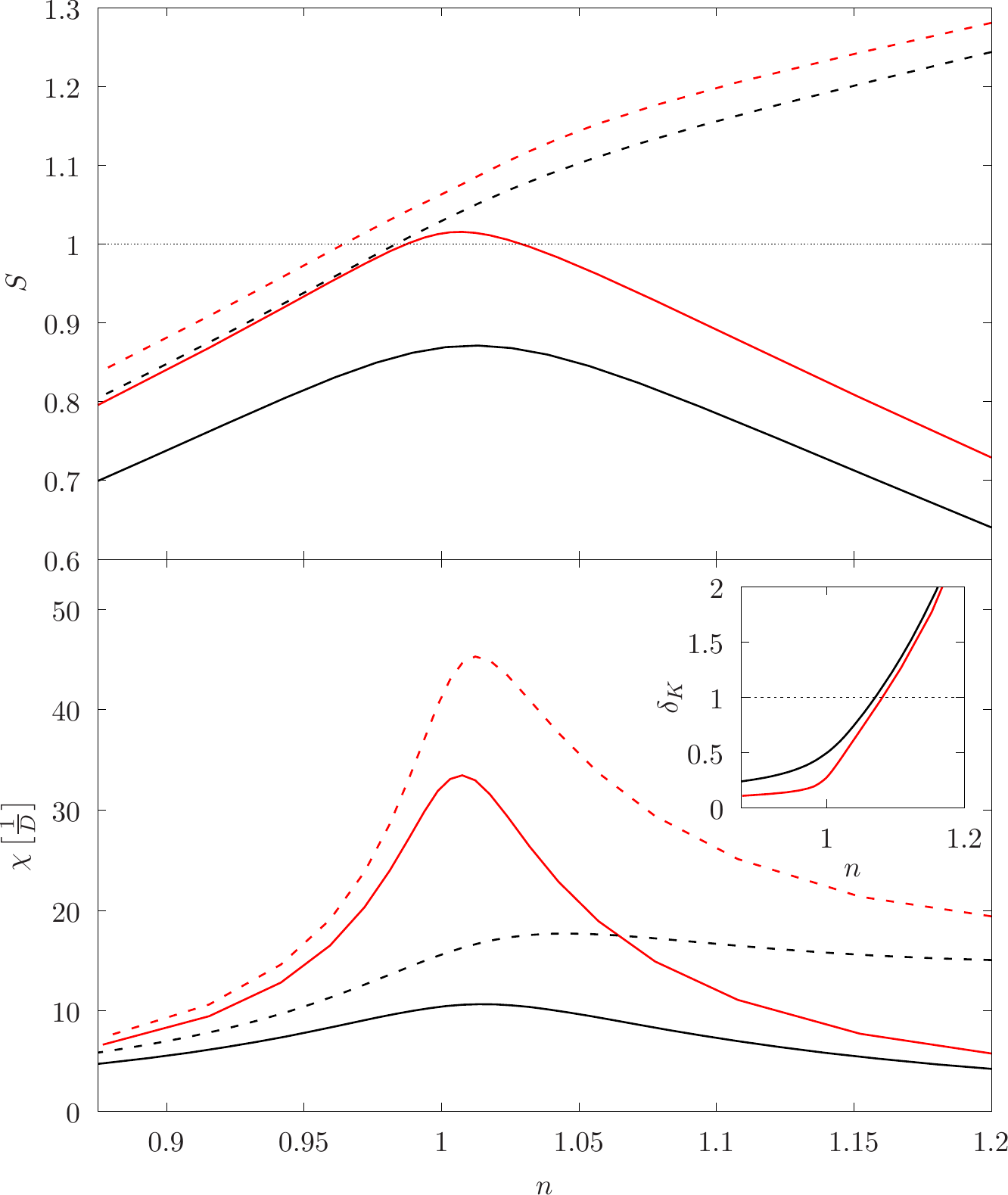}
\caption{(Color online) The upper panel presents the entanglement entropy in the thermodynamic limit for systems with different values of $U_{ab}=5,\,8$ (black and red lines, respectively) and $U_a=U_b=20$.  The results for the entanglement in the spin $S_{\uparrow}$ and pseudospin sectors $S_{\Uparrow}$, for symmetrically coupled devices ($V_a=V_b=0.45$), are presented using dashed and solid lines, respectively. The lower panel presents the susceptibilities for the same set of parameters. The SU(4) symmetry breaking, as measured by the ratio $\delta_K$ (see Sec. \ref{sec:sus}) presented in the inset, increases with increasing $n$ and decreases with increasing $U_{ab}$.
}
\label{fig:U12}
\end{figure}
Figure \ref{fig:U12} presents results for the entanglement entropy in systems with $U_{ab}< U_a=U_b=U<\infty$. The entanglement in the spin partition is larger than in the pseudospin one for all values of the DQD occupancy. While $S_\Uparrow(n)$ has a maximum at $n\simeq 1$ and is nearly symmetric $S_\Uparrow(n)\sim S_\Uparrow(2-n)$ around this point, $S_\uparrow(n)$ increases monotonically with $n$.
The most important observed feature is that while for $n\leq 1$ the symmetry breaking is suppressed for large enough values of $U_{ab}$, for $n>1$ the asymmetry in the entanglement entropy persists and increases with $n$. As a consequence, in order to observe properties associated with an SU(4)-symmetric GS, the DQD occupancy would need to be smaller than $1$ to minimize symmetry breaking effects. 
Interestingly, in the experiments of Keller {\it et al.} \cite{keller2014emergent} the results presented for the conductance as a function of the temperature which are compatible with an SU(4) Kondo behavior are with the gate voltages shifted towards a $n<1$ regime and not in the middle of the Coulomb blockade valley. The conclusions drawn from the analysis of the entanglement entropy are confirmed by the susceptibility results (see lower panel in Fig. \ref{fig:U12}) which show a much smaller SU(4) symmetry breaking, as measured by the ratio $\delta_K$, for $n< 1$ than for $n>1$. This ratio is also more strongly suppressed with increasing $U_{ab}$ for $n\lesssim 1$.

\subsection{Broken hybridization symmetry ($V_a\neq V_b$) }
When the hybridization symmetry is broken ($V_a\neq V_b$) the diagonal energies $\varepsilon_\ell$ need to be adjusted in order to recover a symmetric occupancy in the two QDs, which is a necessary condition to obtain an SU(4)-symmetric GS. As we will see below, this is not however a sufficient condition even for symmetric interactions ($U_a=U_b=U_{ab}=U$). To simplify the discussion and focus of the effect of an hybridization asymmetry we consider in what follows symmetric interactions. The main conclusions do however apply for a more general situation with $U_a,U_b>U_{ab}$.

\begin{figure}[t]
\includegraphics[width=\columnwidth]{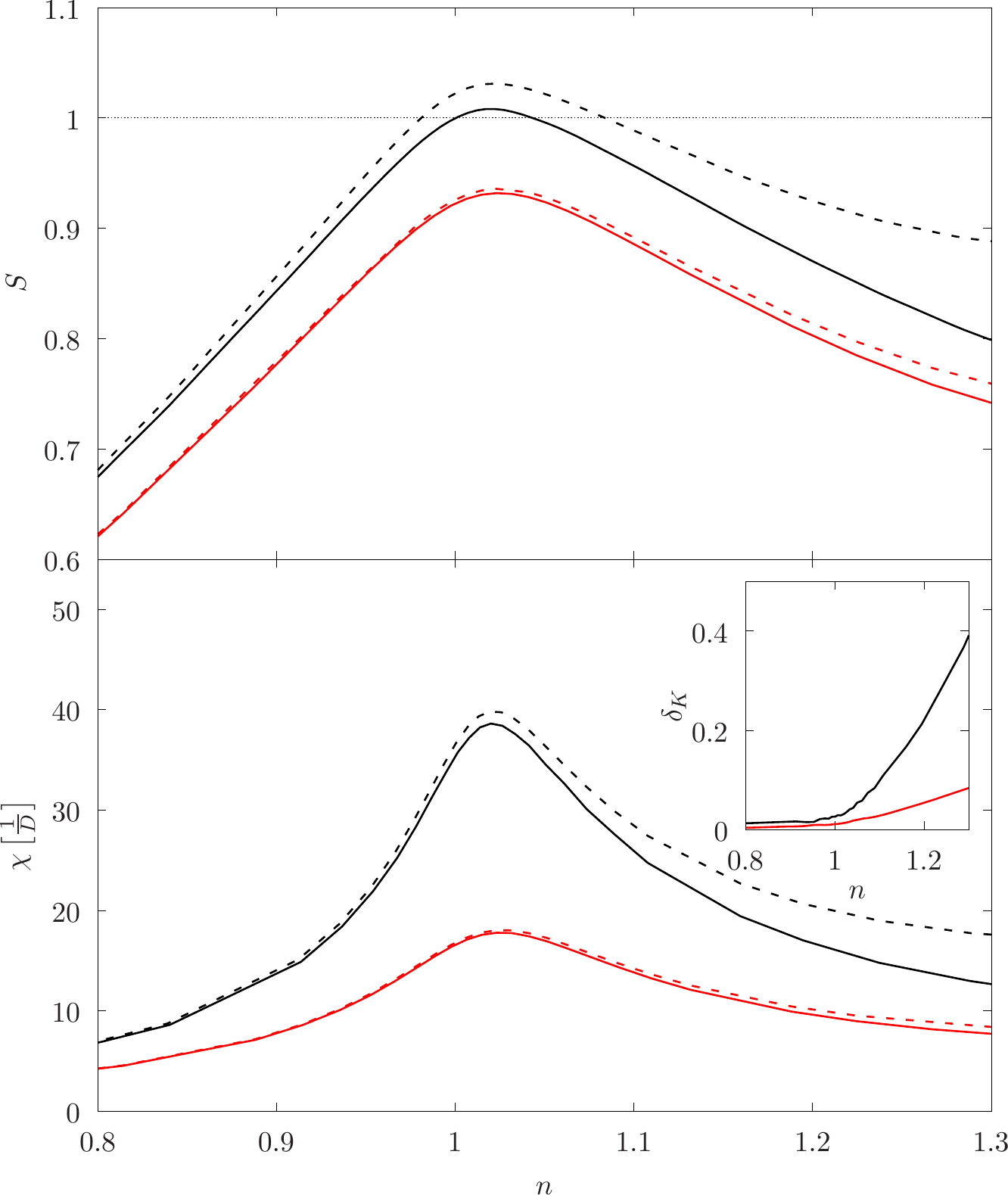}
\caption{(Color online) Entanglement entropy in the thermodynamic limit for systems with SU(4)-symmetric interactions ($U=5$), $V_a=0.45$,  and different values of $V_b=0.25,\,0.35$ (black and red lines, respectively).  The results for the entanglement in the spin $S_{\uparrow}$ and pseudospin sectors $S_{\Uparrow}$, are presented with dashed and solid lines, respectively. The lower panel presents the susceptibilities for the same set of parameters. The SU(4) symmetry breaking, as measured by the ratio $\delta_K$ (see Sec. \ref{sec:sus}) presented in the inset, increases with increasing $n$.
}
\label{fig:diffhyb}
\end{figure}

For the toy model with $U\to \infty$, and fixed $\varepsilon_b$, $V_a$ and $V_b$ it is always possible to obtain an $\varepsilon_a$ such that the two QDs are equally charged (see Appendix). The GS wave function  has the same form as in the case $V_a=V_b$ (see Eq. (\ref{eq:GSUinf})) and is SU(4)-symmetric.
For a finite but large $U\gg V_a,V_b$, a simple $1/U$ first order perturbation theory analysis of the toy model shows that the double occupancy probabilities for QD $a$, QD $b$, and for a single electron on each QD are, respectively, $\sim c V_a^2/U^2$, $\sim c V_b^2/U^2$, and  $c (V_a+V_b)^2/4U^2$. This breaking of the symmetry of the double occupied states in the ground state wave-function, breaks the spin-pseudospin symmetry. This can be observed in the entanglement entropies in Fig. \ref{fig:diffhyb}, calculated using DMRG, where the difference between $S_\uparrow$ and $S_\Uparrow$ increases with increasing $n$. For $n\leq 1$, however, the double occupancy probability is strongly suppressed and the entanglement asymmetry is smaller.

\section{Summary and Conclusions} \label{sec:concl}
We explored the possibility of obtaining SU(4) Kondo behavior in DQD devices where the interactions and the hybridizations are in general non-symmetric. 
To that aim, we analyzed the symmetry of the entanglement entropy for the ground state of a model Anderson Hamiltonian in a wide regime of parameters. 
We obtained both analytical results for a toy model in the narrow band limit and for the full model using numerically exact DMRG calculations.

We find that for non-symmetric interactions or hybridizations, the charge fluctuations to doubly occupied states need to be strongly suppressed in order to avoid breaking the SU(4) symmetry. 
While it has been shown that for a non-SU(4)-symmetric Kondo model the low energy properties of the system are described by an SU(4)-symmetric Kondo Hamiltonian, real systems do present charge fluctuations which can change this picture. 
We find that the double occupancy probability is set primarily by the bare model parameters and essentially unaffected by the Kondo correlations. As the double occupancy probability increases so does the spin-pseudospin symmetry breaking in the ground state wave-function. 
This is nicely illustrated by the toy model which, as such, does not present a Kondo-like renormalization of the parameters, but quantitatively reproduces both the behavior of the charge fluctuations and of the entanglement entropy in the thermodynamic limit.  

We have also calculated the spin and pseudospin susceptibilities. The results are consistent with the entanglement entropy analysis, namely that the SU(4) symmetry breaking can be reduced by suppressing the DQD double occupancy probability. The results suggests that by appropriately setting the parameters a regime with a finite range of temperatures where an SU(4) Kondo scaling could be obtained.

Although the SU(4) symmetry of the ground state wave function is in general broken, the participation of double occupied states, and the degree of symmetry breaking, can be reduced by shifting the level energies of the QDs (experimentally using gate voltages) in such a way that the average charge on the DQD is smaller than $1$. Even in systems where the interactions are the largest energy scale, the level energies would need to be set in a regime such that the DQD average occupancy is $n\lesssim 1$, to suppress significantly the symmetry breaking. 

It is interesting to point out that in the experiments of Keller {\it et al.} the results for the conductance as a function of the temperature, which show a scaling behavior compatible with an SU(4)-symmetric Kondo entanglement, are in a regime where the occupancy of the DQD is expected to be smaller than $1$. A detailed analysis of the conductance as a function of the temperature for different regimes of DQD occupancy would be in order to determine how the degree of SU(4) symmetry breaking in the ground state wavefunction affects the conductance scaling.


\acknowledgments
We acknowledge enlightening discussions with C. A. Balseiro and I. Cirac. This work was partially supported by CONICET PIP0832, SeCTyP-UNCuyo 06C347, and PICT 2012-1069.
\bibliographystyle{apsrev4-1}
\bibliography{references}

\appendix
\section{Toy model (narrow band limit)}
We present here a more detailed analysis of the toy model described in the main text.
This simplified model takes into account a single site for each reservoir [i.e. $L=1$ in Eq. (\ref{eq:tbchain})] which can be interpreted as the narrow conduction band limit \cite{alascio1980localised,allub2015hybrid} of the full model ($L\to \infty$).

As already pointed out, the ground state of this model can be obtained analytically for a restricted, but physically relevant, set of parameters. We first consider the $U_a,U_b \to \infty$ case where the double occupation on either QD is suppressed. The Hamiltonian matrix for the ground state subspace reads
\[ H_S=\left( \begin{array}{cccc}
0 & \sqrt{2}V_a & \sqrt{2}V_b & 0\\
\sqrt{2}V_a & \epsilon_a & 0 & \sqrt{2}V_b\\
\sqrt{2}V_b & 0 & \epsilon_b & \sqrt{2}V_a\\
0 & \sqrt{2}V_b & \sqrt{2}V_a & \epsilon_a+\epsilon_b+U_{ab}\\
\end{array} \right),\]
in the $\{ |0_{ab}\rangle,\, |S_a\rangle,\, |S_b\rangle,\,|S_{ab}\rangle\}$ basis.
Here the state $|0_{ab}\rangle$ corresponds to the DQD devoid of electrons and four electrons located at the reservoir sites, $|S_i\rangle$ corresponds to a singlet state between the QD $i$ and the reservoir site coupled to it, and no electrons in the remaining QD (with the associated reservoir site doubly-occupied), and $|S_{ab}\rangle$ is a direct product of two singlet states formed between each QD and its associated electron bath site. Table \ref{tab:DQDstates} presents a description of these states which are in a charge sector with 4 electrons in the system and a zero total spin projection along the $z$ axis.
\begin{table}[b!]
\centering
    \begin{tabular}{ | l | p{7cm}|}
    \hline
    \textbf{State} & \multicolumn{1}{|c|}{\textbf{Description}} \\ \hline
    $|0_{ab}\rangle$ & $|0\rangle_a \otimes |0\rangle_b = |\underbrace{\uparrow\downarrow}_\text{1a},\underbrace{ - }_\text{a},\underbrace{ - }_\text{b},\underbrace{\uparrow\downarrow}_\text{1b}\rangle$. \\ \hline
    $|S_a\rangle$ & $|S\rangle_a \otimes |0\rangle_b = \frac{1}{\sqrt{2}}\big(|\uparrow,\downarrow, - ,\uparrow\downarrow\rangle - |\downarrow,\uparrow, - ,\uparrow\downarrow\rangle\big)$ \\ \hline
    $|S_b\rangle$ & $|0\rangle_a \otimes |S\rangle_b = \frac{1}{\sqrt{2}}\big(|\uparrow\downarrow, - , \uparrow, \downarrow\rangle - |\uparrow\downarrow, - , \downarrow, \uparrow\rangle\big)$ \\ \hline
    $|S_{ab}\rangle$ & $|S\rangle_a \otimes |S\rangle_b = \frac{1}{2}\big(|\uparrow,\downarrow,\uparrow,\downarrow\rangle - |\uparrow,\downarrow, \downarrow ,\uparrow\rangle +\newline + |\downarrow,\uparrow,\downarrow,\uparrow\rangle - |\downarrow,\uparrow,\uparrow,\downarrow\rangle\big)$ \\ \hline
    $|2_{a}\rangle$ & $|2\rangle_a \otimes |0\rangle_b = |\uparrow\downarrow,-,\uparrow\downarrow,-\rangle$ \\ \hline
    $|2_{b}\rangle$ & $|0\rangle_a \otimes |2\rangle_b = |-,\uparrow\downarrow,-,\uparrow\downarrow\rangle$ \\ \hline
\end{tabular}
\label{tab:DQDstates}
\caption{Notation for the relevant states for the toy model (see Fig. \ref{fig:chain}). }
\end{table}
In what follows, we consider that the site energies ($\varepsilon_a$ and $\varepsilon_b$) are tuned to balance the occupancy of the QDs in the ground state  $|\Psi_{GS}\rangle$ (i.e. $\langle \Psi_{GS}| \hat{n}_a|\Psi_{GS}\rangle=\langle \Psi_{GS}| \hat{n}_b|\Psi_{GS}\rangle$). In the $U_{ab}\to \infty$ limit this can be done setting:
\begin{equation}
\epsilon_a = \frac{\epsilon_b}{2}\bigg(1+\frac{V_a}{V_b}\bigg)+\frac{\sqrt{8(V_a^2+V_b^2)+\epsilon_b^2}}{2}\bigg(1-\frac{V_a}{V_b}\bigg),
\label{eq:param_rel}
\end{equation}
while for finite $U_{ab}$ the condition for charge balance can be numerically obtained.
The ground state wave function in the $U_a,U_b\to \infty$ limit and for a balanced occupancy in the QDs has the form:
\begin{equation}
|\Psi_{\text{GS}}\rangle = \alpha|0_{ab}\rangle + \frac{\beta}{\sqrt{2}}\big(|S_a\rangle + |S_b\rangle\big) + \gamma|S_{ab}\rangle. 
\label{eq:ToyModel_gs}
\end{equation}
The entanglement entropy for partition $\mathcal{H}_\Uparrow\otimes \mathcal{H}_\Downarrow$ can be readily calculated as the system can be regarded as composed by two two-level systems \cite{wootters1998concurrence}
\begin{equation}
S_{\Uparrow}=h\bigg(\frac{1}{2}+\frac{1}{2}\sqrt{1-C^2}\bigg),
\end{equation}
where $h(x)=-x\log _2x-(1-x)\log _2(1-x)$ is Shannon binary entropy and $C$ is Wootters concurrence for the state $|\Psi_{\text{GS}}\rangle$, which equals $|\beta^2-2\alpha\gamma|$.
The entanglement entropy for the partition $\mathcal{H}_\uparrow\otimes \mathcal{H}_\downarrow$, is more complicated as the dimension of the subspace associated with each spin is no longer 2.
The reduced density matrix for the spin partition, $\rho_{\uparrow}:=\text{Tr}_\downarrow|\Psi_{GS}\rangle\langle\Psi_{GS}|$ reads
\[\rho_{\uparrow}=\left( \begin{array}{cccc}
    \frac{\beta^2}{2}+\frac{\gamma^2}{4} & -\frac{\beta}{\sqrt{2}}(\alpha+\frac{\gamma}{2}) & -\frac{\beta\gamma}{2\sqrt{2}} & 0\\
    -\frac{\beta}{\sqrt{2}}(\alpha+\frac{\gamma}{2}) & 1-\frac{\beta^2}{2}-\frac{3\gamma^2}{4} & \frac{\alpha\gamma}{2} & 0\\
-\frac{\beta\gamma}{2\sqrt{2}} & \frac{\alpha\gamma}{2} & \frac{\gamma^2}{4} & 0 \\
0 & 0 & 0 & \frac{\gamma^2}{4} \end{array} \right),\]
in the basis $\{|S_{\uparrow}\rangle,\,|0_{\uparrow}\rangle,\, \,|2_{\uparrow}\rangle,\, |T_{\uparrow}\rangle\}$, which is described in Table \ref{tab:DQDstatesPS}.
\begin{table}[b!]
  \centering
    \begin{tabular}{ | l | c|}
    \hline
\textbf{State} & \multicolumn{1}{|c|}{\textbf{Configuration}} \\ \hline
    $|0_{\uparrow}\rangle$ & $|\underbrace{\Uparrow\Downarrow}_{1\uparrow},\underbrace{ - }_{\uparrow}\rangle$. \\ \hline
    $|S_{\uparrow}\rangle$ & $\frac{1}{\sqrt{2}}\big(|\Uparrow,\Downarrow\rangle -|\Downarrow,\Uparrow\rangle\big)$ \\ \hline
        $|T_{\uparrow}\rangle$ & $\frac{1}{\sqrt{2}}\big(|\Uparrow,\Downarrow\rangle +|\Downarrow,\Uparrow\rangle\big)$ \\ \hline
    $|2_{\uparrow}\rangle$ & $|-,\Uparrow\Downarrow\rangle$. \\ \hline
    \end{tabular}
\caption{Notation for the relevant states for the toy model, for $\uparrow$-spin sector, written in terms of pseudospin degrees of freedom. Notation for $\downarrow$-spin sector is analogous.}
\label{tab:DQDstatesPS}
\end{table}
No simple expression could be obtained in this case for the entropy of entanglement in terms of the GS wave function coefficients.

\begin{figure}[t!]
\includegraphics[width=\columnwidth]{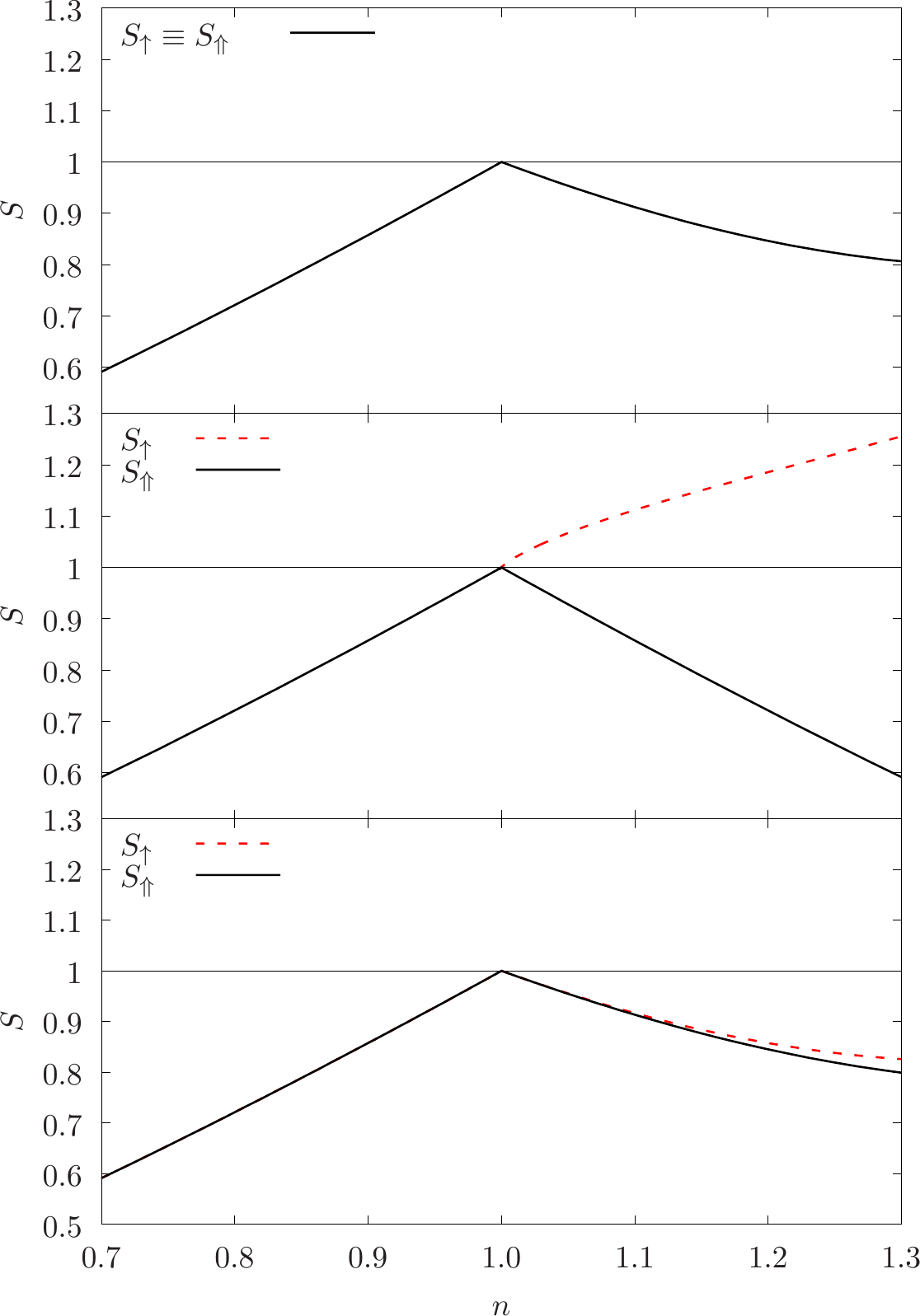}
\caption{(Color online) Entropy of entanglement in the $U\to \infty$ limit for the toy model. a) Symmetric model ($U_a=U_b=U_{ab}=U\to \infty$). b) Asymmetric interactions $U_{ab}<U_a,U_b$. c) Asymmetric hybridizations $V_a=0.45$ and $V_b=0.25$.}
\label{fig:toyentr}
\end{figure}
Figure \ref{fig:toyentr} presents the entanglement entropy for the two partitions considered in the limit of large interaction. For symmetric interactions ($U_a=U_b=U_{ab}=U\to\infty$) and hybridizations ($V_a=V_b$), we trivially have $S_\uparrow\equiv S_\Uparrow$. The entanglement entropy is asymmetric with respect to the $n=1$ axis \footnote{To obtain occupations larger that $1$, the limit $\epsilon_\ell \to -\infty$ needs to be taken together with the $U\to \infty$ limit.} where it presents a cusp [see Eq. (\ref{eq:entrn1})]. For finite values of $U$ (not shown) the cusp is rounded and $S(n)$ presents a maximum at $n\sim 1$ as in the $L\to \infty$ case. 
In the large interaction limit, the entropy preserves the spin-pseudospin symmetry for $n\leq 1$ when the symmetry of the parameters is broken $U_a,U_b>U_{ab}$ [see Fig.\ref{fig:toyentr}a)] or $V_a\neq V_b$ [see Fig. \ref{fig:toyentr}b)]. This can be readily seen analyzing the entanglement entropy of the ground state wave function of Eq. (\ref{eq:ToyModel_gs}) when the double occupation is suppressed and $\gamma=0$. For $n>1$ in the same limit, the different doubly occupied states have a different amplitude in the ground state wave function which breaks the spin-pseudospin symmetry.

In order to analyze the symmetry of the ground state wave function it is instructive to consider what happens to the states when transformed according to a map $\hat{P}$ which exchanges spin and pseudospin degrees of freedom, i.e. $\hat{P}:(\uparrow,\downarrow)\leftrightarrow(\Uparrow,\Downarrow)$. We detail the procedure for the singlet state in QD $a$ ($\Uparrow$), $|S_a\rangle$:

\begin{align}
\hat{P}|S_a\rangle=&\hat{P}\frac{1}{\sqrt{2}}\big(c_{\uparrow\Uparrow}^{\dagger}d_{\downarrow\Uparrow}^{\dagger}-c_{\downarrow\Uparrow}^{\dagger}d_{\uparrow\Uparrow}^{\dagger}\big)c_{\uparrow\Downarrow}^{\dagger}c_{\downarrow\Downarrow}^{\dagger}|\text{vac}\rangle\nonumber\\
=&\frac{1}{\sqrt{2}}\big(c_{\uparrow\Uparrow}^{\dagger}d_{\uparrow\Downarrow}^{\dagger}-c_{\uparrow\Downarrow}^{\dagger}d_{\uparrow\Uparrow}^{\dagger}\big)c_{\downarrow\Uparrow}^{\dagger}c_{\downarrow\Downarrow}^{\dagger}|\text{vac}\rangle\nonumber\\
=&\frac{1}{\sqrt{2}}\big(-c_{\uparrow\Uparrow}^{\dagger}c_{\downarrow\Uparrow}^{\dagger}d_{\uparrow\Downarrow}^{\dagger}c_{\downarrow\Downarrow}^{\dagger}+c_{\downarrow\Uparrow}^{\dagger}d_{\uparrow\Uparrow}^{\dagger}c_{\uparrow\Downarrow}^{\dagger}c_{\downarrow\Downarrow}^{\dagger}\big)|\text{vac}\rangle\nonumber\\
=&\frac{1}{\sqrt{2}}\big(|\uparrow\downarrow,-,\uparrow,\downarrow\rangle-|\downarrow,\uparrow,-,\uparrow\downarrow\rangle\big)\nonumber
\end{align}

\noindent We use $c^{\dagger}$ to denote electron-creation operators for the bath sites, dropping the subindex 0 as there is only one for each QD. In the second step above, we reorder creation operators so as to have all $\Uparrow$-ones to the left. Performing a similar operation on each of the other relevant states, we get:

\begin{align}
\hat{P}|0_{ab}\rangle=&-|0_{ab}\rangle\nonumber\\
\hat{P}|S_{b}\rangle=&\frac{1}{\sqrt{2}}\big(|\uparrow\downarrow,-,\downarrow,\uparrow\rangle-|\uparrow,\downarrow,-,\uparrow\downarrow\rangle\big)\nonumber\\
\hat{P}|S_{ab}\rangle=&-\frac{1}{2}\big(|\uparrow,\downarrow,\uparrow,\downarrow\rangle + |\downarrow,\uparrow,\downarrow,\uparrow\rangle\nonumber\\ &+ |-,\uparrow\downarrow,-,\uparrow\downarrow\rangle+|\uparrow\downarrow,-,\uparrow\downarrow,-\rangle\big)\nonumber
\end{align}

The state with no electrons in the DQD is trivially invariant (it only changes sign), whereas $|S_a\rangle$, $|S_b\rangle$ and $|S_{ab}\rangle$ are not. However, it is straightforward to realize that the combination $|S_a\rangle+|S_b\rangle$ is indeed invariant: $\hat{P}\big(|S_a\rangle+|S_b\rangle\big) = -\big(|S_a\rangle+|S_b\rangle\big)$. 
As a consequence, a finite amplitude of the state $|S_{ab}\rangle$ in the ground state will lead to a spin-pseudospin symmetry breaking, unless the states with a double occupancy on a single QD have the same amplitude.

\begin{figure}[ht]
\includegraphics[width=\columnwidth]{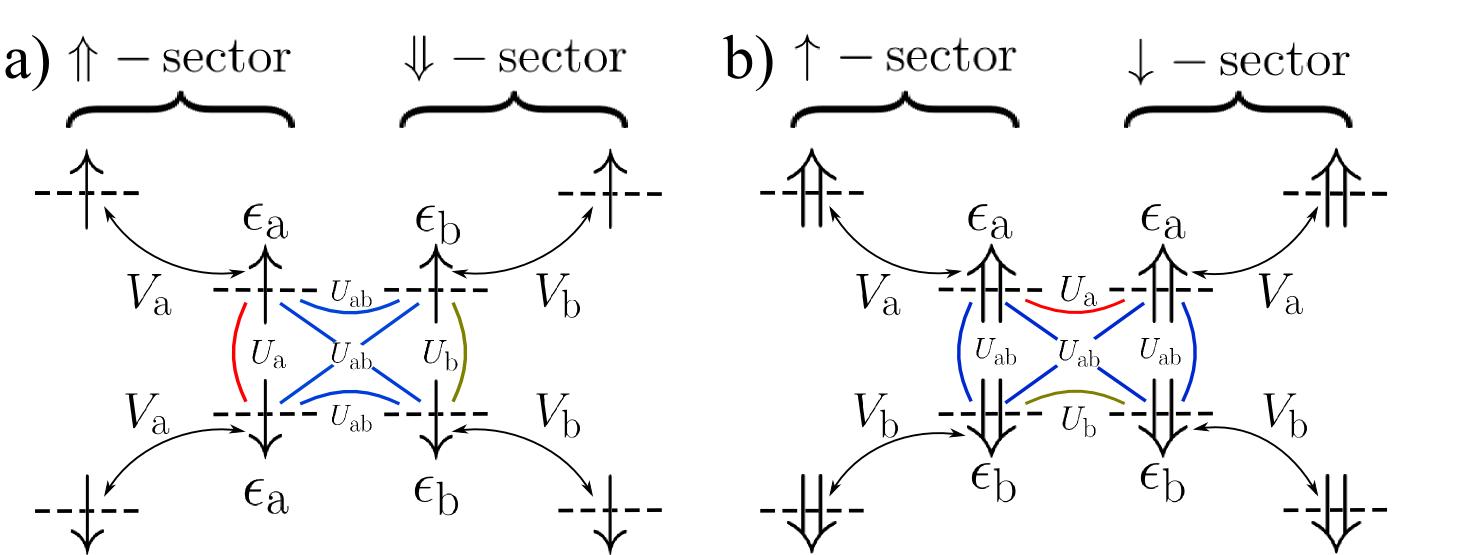}
\caption{(Color online): schemes showing the different interaction channels for the electrons in the DQD system. The outermost levels represent bath electrons, while innermost ones represent DQD electrons.}
\label{fig:mapping}
\end{figure}

Finally, in Fig. \ref{fig:mapping}, we present the interaction and coupling channels of the DQD for a partition in spin (Fig. \ref{fig:mapping} (a)) and pseudospin (Fig. \ref{fig:mapping} (b)) sectors. In the case where $U_i>U_{ab}$, electrons in the DQD can be thought as interacting through a symmetric coulombian interaction $U_{ab}$, plus an antiferromagnetic-like term between the pseudospin degrees of freedom with coupling constant $U_i-U_{ab}$, which increases correlations and thus entanglement for the $\mathcal{H}_\uparrow\otimes \mathcal{H}_\downarrow$ partition.

\end{document}